\documentclass{aastex}
\usepackage{spr-astr-addons}
\usepackage{url}
\usepackage{diagbox}
\usepackage{graphicx}
\usepackage{sidecap}

\def\simlt{\lower.5ex\hbox{$\; \buildrel < \over \sim \;$}}
\def\simgt{\lower.5ex\hbox{$\; \buildrel > \over \sim \;$}}
\def\kms{km~s$^{-1}$}
\def\Kkms{K~km~s$^{-1}$}
\def\vlsr{v_{\rm LSR}}
\def\schi{{\sc H\,i}}
\def\schii{{\sc H\,ii}}
\def\h2{{\rm H$_2$}}
\def\hco{{\rm HCO$^+$}}
\def\alphaJ{{$\alpha_{2000}$}}
\def\deltaJ{{$\delta_{2000}$}}
\def\Av{\rm A_v}
\def\ebv{\rm E(B\!-\!V)}
\def\whi{W_{\rm HI}}
\def\wco{W_{\rm CO}}
\def\ntot{{N({\rm H})}}
\def\nhi{{N({\rm HI})}}
\def\nh2{{N({\rm {H_2}})}}
\def\nhco{{N({\rm HCO^+})}}
\def\nhcoth2{{N{\rm(H_2)}_{\rm HCO^+}}}
\def\ncoth2{{N{\rm(H_2)}_{\rm CO}}}

\def\lab{{LAB}}

\begin{document}

\title{Galactic HCO$^+$ Absorption toward Compact Extragalactic Radio Sources}
\slugcomment{}
\shorttitle{Galactic HCO$^+$ Absorption toward Compact Extragalactic Radio Sources}
\shortauthors{G. Park et al.}

\author{Geumsook Park\altaffilmark{\dag,1,2}} 
\author{Bon-Chul Koo\altaffilmark{2}}   
\author{Kee-Tae Kim\altaffilmark{1}} 
\author{Do-Young Byun\altaffilmark{1,3,4}}  
\and
\author{Carl E. Heiles\altaffilmark{5}}

\altaffiltext{1}{Korea Astronomy and Space Science Institute, 
Daedeokdae-ro, Yuseong-gu, Daejeon 34055, Republic of Korea}
\altaffiltext{2}{Department of Physics and Astronomy, Seoul National University, 
1 Gwanak-ro, Gwanak-gu, Seoul 08826, Republic of Korea}
\altaffiltext{3}{Korea University of Science and Technology, 
217 Gajeong-ro, Yuseong-gu, Daejeon 34113, Republic of Korea}
\altaffiltext{4}{Yonsei University, Yonsei-ro 50, Seodaemun-gu, Seoul 03722, Republic of Korea}
\altaffiltext{5}{Radio Astronomy Lab, UC Berkeley 
601 Campbell Hall, Berkeley, CA 94720, USA}
\altaffiltext{\dag}{pgs@kasi.re.kr}

\begin{abstract}
As part of the search for the ``dark molecular gas (DMG),''
we report on the results of \hco~$J=1-0$ absorption observations 
toward nine bright extragalactic millimeter wave continuum sources.
The extragalactic sources are at high Galactic latitudes ($|b| > 10\arcdeg$) 
and seen at small extinction ($\ebv \lesssim 0.1$~mag).
We have detected the \hco\ absorption lines toward two sources, B0838+133 and B2251+158. 
The absorption toward B2251+158 was previously reported, 
while the absorption toward B0838+133 is a new detection.
We derive hydrogen column densities or their upper limits 
toward the nine sources from our observations
and compare them to those expected from CO line emission and far-infrared dust continuum emission.
Toward the seven sources with no \hco\ detection,
CO emission has not been detected, either.  
Thus the sight lines are likely to be filled with almost pure atomic gas.
Toward the two sources with \hco\ detection,
CO emission has been also detected.
Comparison of the \h2\ column densities from \hco\ absorption and CO emission
suggests a non-negligible amount of DMG toward B0838+133.
\end{abstract}



\section{Introduction}
\label{sec:intro}

The Interstellar Medium (ISM) is mainly composed of hydrogen in three phases: 
atomic (\schi), molecular (\h2), and ionized (\schii). 
\schi\ atoms are directly observed in the \schi\ 21-cm line,
while most \h2\ molecules are in so cold states 
that they cannot be excited by any radiative transition.
Instead, carbon monoxide (CO) molecular lines are usually used
to trace \h2.
That is, the amount of CO emission has been used to infer that of molecular gas,
which is almost entirely composed of \h2, 
by using an empirical CO-\h2\ conversion factor.
Recently, however, researchers have discovered ``dark gas,'' 
invisible in \schi\ and CO, in the solar neighborhood;
this ``dark gas'' has a non-negligible mass.
It can be found by excess $\gamma$ ray emission \citep[e.g.,][]{grenier2005,abdo2010}
or excess dust emission \citep[e.g.,][]{planck2011a19,planck2011a24}.
These observational results imply there is an additional ISM component
that cannot be traced by \schi\ or CO line observations.
The ``dark gas'' component is generally considered to be a molecular gas,
a so-called ``dark molecular gas (DMG)''
\citep[e.g.,][]{lucas1996},
despite another suggestion by \citet{fukui2014, fukui2015} 
that optically thick and cold \schi\ gas mainly contributes ``dark gas.''
 
Theoretically, 
the presence of DMG is supported by the photodissociation region (PDR) model
\citep[e.g.,][]{vanDishoeck1988, wolfire2010}.
The PDR model predicts an intermediate layer between \schi-to-\h2\ and \h2-to-CO transitions,
where CO cannot survive UV photodissociation
but \h2\ can self-shield.
\citet{wolfire2010} inferred that the \schi-to-\h2\ transition
is located at a visual extinction of $\Av \simeq 0.2$~mag,
which is consistent with observational findings;
for example,
\citet{paradis2012} and \citet{planck2011a19}
found the threshold to be 0.2~mag and 0.4~mag, respectively.
Corresponding reddenings $\ebv$ are 0.065 and 0.13~mag, respectively,
assuming that $\Av/\ebv = 3.1$ for the diffuse ISM \citep{savage1979}.
The main chemical route associated with CO in diffuse clouds predicts
that OH, C$^{+}$, and \hco\ can be observable before CO formation \citep{vanDishoeck1988}.
Such elements or molecules would be useful tracers for CO-dark molecular gas.
\citet{liszt1996} and \citet{lucas1996} confirmed that
OH and \hco\ do reliably trace DMG.
\citet{tang2017} also showed that C$^{+}$ could be a useful tracer for DMG.

\citet[][hereafter, LL96]{lucas1996} surveyed \hco\ absorption
toward thirty lines-of-sight (LOSs) of extragalactic background continuum sources,
finding detectable absorption lines for eighteen sources.
Since then, there have been several studies of \hco\ absorption lines
\citep[][and see a compilation in Appendix E of \citet{liszt2010}]{liszt2000, liszt2010}.
In this paper, 
using the Korean VLBI Network (KVN) 21~m telescope in the single dish mode,
we present the observational results of
\hco\ absorption lines toward several background sources missing before.
In Sections~\ref{sec:obs} and \ref{sec:res},
we describe our KVN observations and results, respectively.
In Section~\ref{sec:disc},
we discuss gas properties of the individual LOSs.
Section~\ref{sec:sum} summarizes the paper.

\section{Observations}
\label{sec:obs}

Using the KVN 21-m telescope at the Yonsei station in the single dish mode,
we observed nine positions in the transition $J$~=~1--0 of \hco\ (89.188526 GHz)
\citep{kim2011, lee2011}.
The positions lie on a background of extragalactic compact radio sources
(such as quasars or AGN), 
which are listed in Table~\ref{tab:targets};
our intent was to observe absorption lines from Galactic dark molecular gas in the foreground.
The observations toward B0838+133 and B2251+158 were performed on 07 February 2013,
and the others during the period from September 2014 to January 2015.
The digital spectrometer was set to have 4096 channels
with a bandwidth of 64~MHz ($\sim 216$~\kms\ at 89~GHz) 
and centered at $\vlsr = 0$~\kms.\footnote[1]{
Velocities ($\vlsr$) in this paper is 
with respect to the local standard of rest (LSR).}
A single channel width is 0.016~MHz, 
giving a velocity resolution of 0.05~\kms.
The 21-m telescope had a main beam efficiency of $\sim 36\%$
and a beam size (Full Width at Half Maximum; FWHM) of 31\arcsec\ at 89~GHz.
Observations were done in dual polarization mode.
While data for B0838+133 and B2251+158 were taken by position switching (PS),
the other data were obtained by frequency switching (FS).
For the off-position of the PS mode,
four locations, ($-$1\arcmin, $-$1\arcmin), (+1\arcmin, $-$1\arcmin), 
($-$1\arcmin, +1\arcmin), and (+1\arcmin, +1\arcmin)
from the on-position, were alternately observed.
For the FS mode, the frequency offset was set to 16~MHz.
Pointing observations were usually performed 
approximately once every hour ($\sim 2.5$~hours at the longest).
We used only data having system temperature ($T_{sys}$) less than 400~K.
Total exposure time ($t_{tot}$) of the data that were utilized finally  
are noted in the seventh column of Table~\ref{tab:targets}.
Seven sources were missed in previous surveys,
while two sources, B0316+413 and B2251+158, were observed in LL96.

\begin{table*}[ht!]
\scriptsize
\setlength{\tabcolsep}{0.035in}
\caption{Target Sources with Observational Log \label{tab:targets}}
\begin{tabular}{@{}lcccc cccl@{}}
\tableline
Name & Other Name & \alphaJ & \deltaJ &
$\ell$& $b$ & $t_{\rm tot}$ & $T_{\rm A^{*},RMS}$ &
{Observing Dates} \\ 
 &  & (h:m:s) & (d:m:s) & 
(\arcdeg) & (\arcdeg) & (min) & (mK) &
{(YYMMDD)}  \\
(1) & (2) & (3) & (4) & 
(5) & (6) & (7) & (8) & 
{(9)} \\
\tableline
B0316$+$413 & 3C84                      & 03:19:48.16 & $+$41:30:42.1 & 150.576 & $-$13.261 & 345 & ~7 &        141007, 141011, 141201, 150104 \\
B0420$-$014 & PKS 0420-01               & 04:23:15.80 & $-$01:20:33.1 & 195.290 & $-$33.140 & 234 & ~7 &        141201                         \\
B0838$+$133 & 3C207                     & 08:40:47.59 & $+$13:12:23.6 & 212.968 & $+$30.139 &  88 & 10 &        130207                         \\
B0851$+$202 & OJ287                     & 08:54:48.88 & $+$20:06:30.6 & 206.812 & $+$35.821 & 398 & ~7 &        141007, 141009, 141011, 141201 \\
B1228$+$126 & {\tiny VIRGO A, NGC 4486} & 12:30:49.42 & $+$12:23:28.0 & 283.778 & $+$74.491 & 300 & ~7 &        141109, 141116, 141201, 150104 \\
B1633$+$382 & 4C38.41                   & 16:35:15.49 & $+$38:08:04.5 & ~61.086 & $+$42.337 & 435 & ~7 & {\tiny 140930, 141007, 141109, 141116, 141201, 150104} \\
B2223$-$052 & 3C446                     & 22:25:47.26 & $-$04:57:01.4 & ~58.960 & $-$48.843 & 250 & ~7 &        141109, 141116, 141201, 150104 \\
B2249$+$185 & 3C454                     & 22:51:34.74 & $+$18:48:40.1 & ~87.354 & $-$35.648 & 283 & ~8 &        141009, 141011, 141109, 150104 \\
B2251$+$158 & 3C454.3                   & 22:53:57.75 & $+$16:08:53.6 & ~86.111 & $-$38.184 &  52 & 13 &        130207                         \\
\tableline
\end{tabular}
\tablecomments{Columns are as follows: (1) IAU B1950.0 source name; 
              (2) source name in other catalogs;
              (3)--(6) source coordinates; 
              (7) total integration time with  $T_{\rm sys} < 400$~K;
	      (8) RMS noise level on the $T_{\rm A^*}$ scale at a velocity resolution of 0.1~\kms; 
	      (9) Observing dates (2-digit year to be interpreted as 20xx)}
\end{table*}

When planning observations, 
the background galaxies in Table~\ref{tab:targets} were selected as
bright radio continuum sources mostly with flux densities $> 3$~Jy
around the observing frequency,
but about half had lower values or were even invisible during our observing period
because of their flux variability.
The flux densities are 
listed in the second column of Table~\ref{tab:gaussfit}.
Flux measurements are performed with Gaussian fittings of average ``cross-scan'' data 
obtained during our KVN observations.
There was no cross-scan data for B0838+133, 
so we assumed its flux based on data from the nearest dates in the ALMA calibrator database.
For B2249+185, most of the observing dates had no signal.
B2249+185 may have been invisible during the observing season,
which implies that the source may have been radio-quiet during those days.

\begin{table*}
\small
\caption{\hco\ Absorption Line Parameters toward Nine Sources\label{tab:gaussfit}}
\begin{tabular}{@{}lcccccc@{}}
\tableline
Name & S$_{\rm 86\,GHz}$\tablenotemark{a} & $v_{0}$ & $\Delta v_{\rm FWHM}$
& $\tau_{0}$ & $\int\tau dv$ & $\nhco$ \\
& (Jy) & (\kms) & (\kms) &  & (\kms) & ($10^{12}{\rm cm}^{-2}$)\\
\tableline
B0316$+$413 & 19.0$\pm$0.3 & \nodata          & \nodata       &$<$~0.02       &$<$~0.02       &$<$~0.02        \\
B0420$-$014 & ~1.2$\pm$0.1 & \nodata          & \nodata       &$<$~0.28       &$<$~0.30       &$<$~0.33        \\
B0838$+$133 & $\sim0.7$    & $+$4.44$\pm$0.04 & 0.47$\pm$0.09 & 1.44$\pm$0.23 & 0.72$\pm$0.18 & 0.81$\pm$0.20  \\
B0851$+$202 & ~4.0$\pm$0.1 & \nodata          & \nodata       &$<$~0.09       &$<$~0.10       &$<$~0.11        \\
B1228$+$126 & ~4.5$\pm$0.1 & \nodata          & \nodata       &$<$~0.08       &$<$~0.09       &$<$~0.10        \\
B1633$+$382 & ~1.7$\pm$0.1 & \nodata          & \nodata       &$<$~0.21       &$<$~0.22       &$<$~0.25        \\
B2223$-$052 & ~0.9$\pm$0.1 & \nodata          & \nodata       &$<$~0.42       &$<$~0.45       &$<$~0.50        \\
B2249$+$185 & $< 0.3$      & \nodata          & \nodata       & \nodata       & \nodata       & \nodata        \\
B2251$+$158 & ~5.0$\pm$0.1 & $-$9.47$\pm$0.09 & 0.92$\pm$0.24 & 0.22$\pm$0.03 & 0.22$\pm$0.06 & 0.24$\pm$0.07  \\
            &              &$-$10.84$\pm$0.33\phn & 1.27$\pm$0.63 & 0.07$\pm$0.02 & 0.09$\pm$0.06 & 0.11$\pm$0.06  \\
\tableline
\end{tabular}
\tablenotetext{a}{The flux densities of the background sources except B0838$+$133
		  were measured using the KVN 21-m telescope during our observation period.
                  The conversion factor from ${T_{\rm A^*}}$ to flux density 
                  is $\sim16$~Jy~K$^{-1}$.
		  The 86~GHz flux of B0838$+$133 assumes 0.7~Jy, 
		  which includes one valid decimal place, based on a known database:
		  0.73~Jy at 91.5~GHz on 2013--12--29 in the ALMA calibrator database (https://almascience.eso.org/sc/).
                  }
\end{table*}

\section{Results}
\label{sec:res}

\begin{SCfigure*}[][ht!]
\includegraphics[width=129mm]{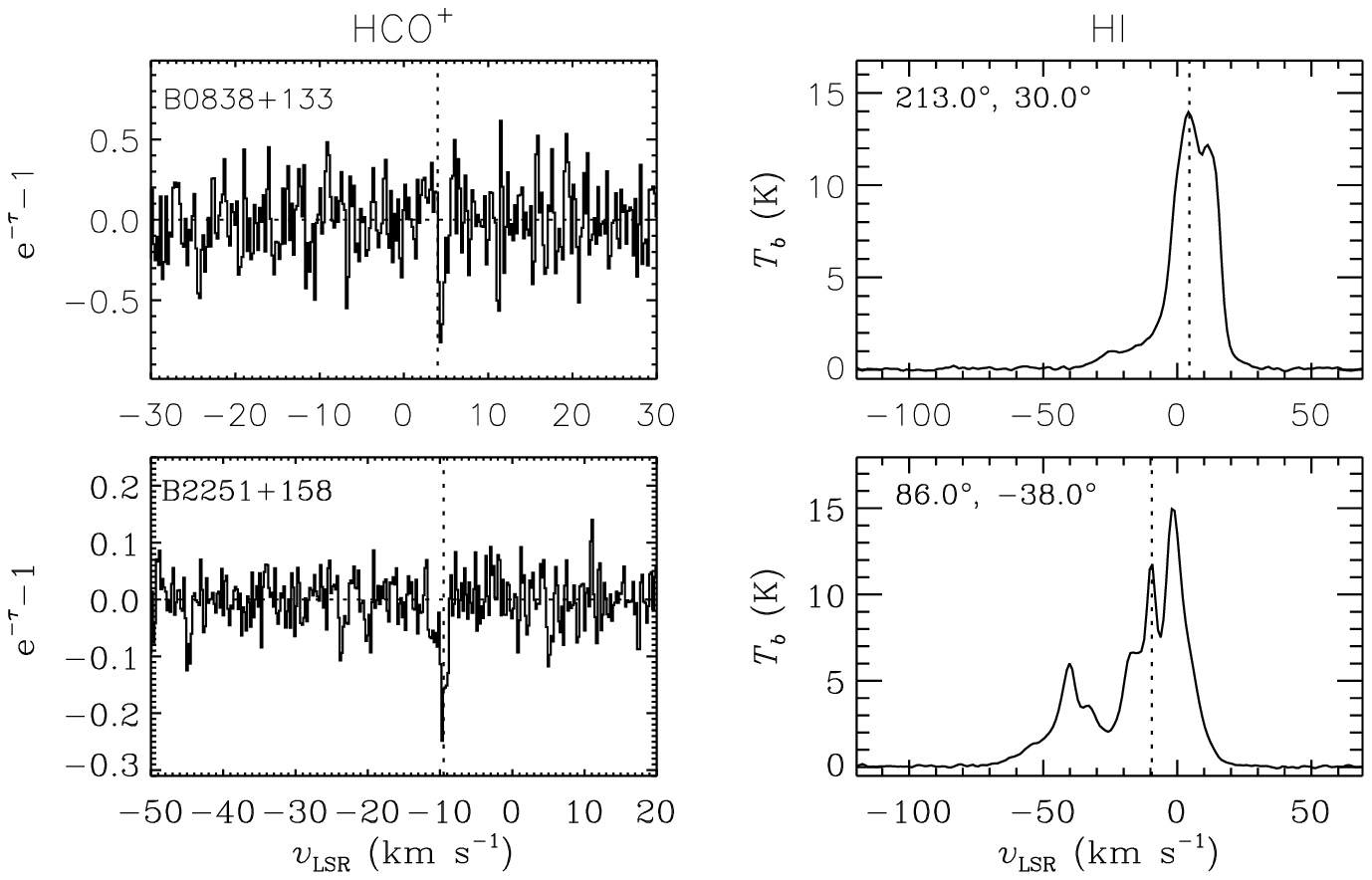}
\caption{
Left: \hco\ absorption line profiles toward B0838+133 and B2251+158.
Hanning smoothing was applied twice, resulting in a velocity resolution of 0.2~\kms.
The vertical dotted line marks the adjacent \schi\ peak velocity at which \hco\ absorption line is detected.
Right: \schi\ line profile toward the two sources.
The profiles are from the LAB survey (FWHM = 30\arcmin) at Galactic coordinates
written at the upper left corner of each panel.
The vertical dotted line marks where \hco\ absorption line is detected.
}\label{fig:spec_det}
\end{SCfigure*}

\begin{figure}[ht!]
\includegraphics[width=84mm]{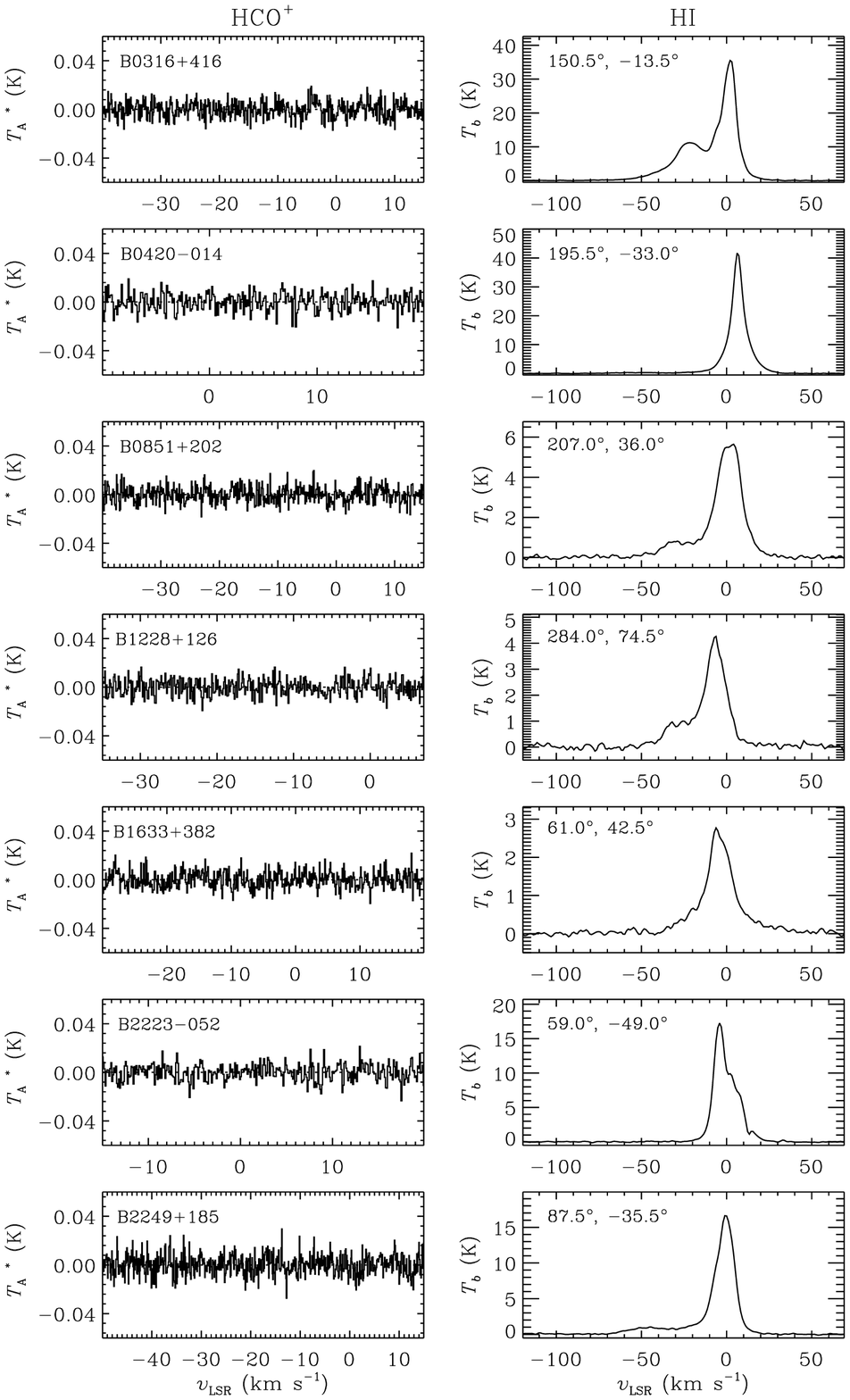}
\caption{
Same as Figure~\ref{fig:spec_det} but for \hco-undetected sources. 
}\label{fig:spec_nondet}
\end{figure}

Figs.~\ref{fig:spec_det} and \ref{fig:spec_nondet} show
the spectra observed for the \hco-detected and -undetected sources, respectively.
For the \hco\ spectra,
Hanning smoothing is applied once or twice using `CLASS' 
from the GILDAS software package\footnote[2]{\url{http://www.iram.fr/IRAMFR/GILDAS}}. 
Then, each spectrum is baseline-corrected by $n$th-order polynomial fitting:
third and first for B0838+133 and B2251+158, respectively,
and seventh (or fifth) for the others.
The spectral velocity range shown in Fig.~\ref{fig:spec_nondet}
is based on where the Galactic \schi\ emission of
the Leiden/Argentine/Bonn (\lab) all-sky survey data
\citep[$0\fdg5$-pixel with an angular resolution of $\sim 36\arcmin$;][]{kalberla2005} 
is seen in the same LOS;
the \schi\ line profiles toward our nine LOSs are displayed
together in Figs.~\ref{fig:spec_det} and \ref{fig:spec_nondet}.
The resulting root-mean-square (RMS) antenna temperature values
at a velocity resolution of 0.1~\kms\
are listed in the last column of Table~\ref{tab:targets};
typical RMS noise level ($1\sigma$) is 6~mK.
We detected an \hco\ absorption line in two sources:
the existence of an \hco\ absorption line in the LOS of B2251+158
has already been reported by LL96,
while we detected for the first time an absorption feature toward B0838+133.
However, none of the other samples show any absorption lines.
Interestingly, a weak blue wing is seen in the \hco\ line of B2251+158,
as noted in \citet{liszt2012}.

For the two detected cases,
we applied a Gaussian fit
with an assumption of a single component for B0838+133 and two components for B2251+158.
This was done because the latter's profile shows one more negative-velocity component 
that is weak but likely real;
this component also appears 
in profiles from previous observations \citep[LL96;][]{liszt2000,liszt2012}.
Table~\ref{tab:gaussfit} presents the resultant parameters for 
the central velocity ($v_0$), velocity width ($\Delta v_{\rm FWHM}$), optical depth ($\tau$) 
at $v_0$, and integrated optical depth.
Uncertainties of the first three parameters were taken from
those derived during Gaussian fit (GAUSSFIT in IDL);
the last one was from the results of Monte Carlo simulations using imaginary profiles
formed from observed spectra with 1$\sigma$ RMS noise. 
For the undetected cases, except for B2249+185, 
we give an upper limit assuming one Gaussian component 
with peak temperature of $3 \times T_{A^{*},{\rm RMS}}$ and a line width of 1~\kms.
For reference,
the mean \hco\ line width of detected sources in LL96 is 0.95~\kms.
Our results are consistent with the results of LL96
for B0316+413 and B2251+158. 
For B2251+158, 
\hco\ profiles were reported in \citet{liszt2000} as well as LL96,
and they gave results of single-component Gaussian fit,
which is consistent 
with the total optical depth of our two components within $1\sigma$ uncertainty.
As shown in the last column in Table~\ref{tab:gaussfit},
we derived the \hco\ column density, $\nhco$, 
using the relationship with the integrated \hco\ optical depth
\citep[e.g., see][]{liszt2010}, i.e.,
\begin{equation}
\nhco = 1.12\times10^{12}~{\rm cm}^{-2}\int\tau_{\rm HCO^+}dv~({\rm km~s^{-1}})^{-1}.
\end{equation}

\section{Discussion}
\label{sec:disc}

We wondered if there is ``dark gas'' indeed toward the \hco-detected LOSs
or no dark gas toward the undetected LOSs. 
To answer this question,
we consider the total column density of hydrogen nuclei, $\ntot$, in the LOS.
$\ntot$ can be determined by the sum of column densities of \schi\ and \h2,
i.e., $\ntot\,=\,\nhi\,+\,2\nh2$, ignoring the ionized gas.
Alternatively, it can be inferred using
the relation with optical reddening $\ebv$.
Since these two methods are independent,
we can discuss the implications of our observational results 
by comparison between measurements of the two approaches.

\subsection{$\nhi$ and $\nh2$ derived from radio tracers}
\label{sec:nh_conv}

As mentioned in Section~\ref{sec:intro},
$\nhi$ is obtained directly by \schi\ 21-cm line observations, 
while $\nh2$ is usually inferred from integrated CO intensity ($\wco$)
using the empirical relationship between $\nh2$ and $\wco$. 
That is, $\nhi$ is calculated using the equation of
\begin{equation}
\nhi/\whi = 1.82\times10^{18}~{\rm cm^{-2}}~({\rm K~km~s^{-1}})^{-1},
\end{equation}
where $\whi = \int{T_{\rm b,\,HI}}\,dv$, and with an assumption of optically thin conditions, 
and $\nh2$ is derived from
\begin{equation}
\nh2/\wco = 2.0\times10^{20}~{\rm H_2}~{\rm cm^{-2}}\,({\rm K~km~s^{-1}})^{-1} 
\end{equation}
with $\pm30$\% uncertainty \citep{bolatto2013}.
As another approach,
$\nh2$ can be measured using the relation with $\nhco$ \citep[e.g.,LL96;][]{liszt2010}, i.e.,
\begin{equation}
\nhco/\nh2 = 3\times10^{-9}. 
\end{equation}
For $\whi$,
we obtained a line profile at a given position from the \lab\ data
(see Figs.~\ref{fig:spec_det} and \ref{fig:spec_nondet})
and integrated it over LSR velocities of $\pm$150~\kms\
wide enough to contain most Galactic \schi\ gas. 
Moreover, $\wco$ values were taken from the literature of 
\citet{liszt1993}, \citet{liszt2010}, and \citet{li2018}.
Table~\ref{tab:otherdata} lists the values of $\whi$ and $\wco$ that we adopted.
In the 2nd-4th columns of Table~\ref{tab:cN} we list
the \schi\ and \h2\ column densities derived using equations~(2)--(4);
the sums of different \h2\ measurements are in the 5th-6th columns.
CO line emission was detected toward the two sources of B0838+133 and B2251+158,
but not in the others except B1228+126, which has no available literature data.
CO emission toward B0838+133 was not detected in the previous survey of \citet{liszt1994},
but was detected in a recent deeper survey of \citet{li2018}. 
Although CO emission is observed toward both \hco-detected sources,
$\wco$ of B2251+158 is about two times larger than that of B0838+133.
For B0838+133,
the \hco\ absorption line is decomposed as a single component
at a velocity similar to that of the CO emission line \citep{li2018},
but $\nh2$ derived from \hco\ is three times larger than
$\nh2$ inferred from CO.
The given $\nhco$ value that was used for $\nh2$ is uncertain,
but it may still be possible
that molecular gas not traced by CO exists toward B0838+133.

\begin{table}[ht!]
\tablewidth{1.0\columnwidth}
\caption{\schi\ and CO Line Intensities and $\ebv$ toward Nine Sources \label{tab:otherdata}}
\begin{tabular}{@{}lccc@{}}
\tableline
Name  & $\whi$\tablenotemark{1}   & $\wco$\tablenotemark{2}  & $\ebv$\tablenotemark{3}  \\ 
  & (\Kkms)   & (\Kkms)  & (mag)   \\
\tableline
  B0316$+$413 &  710.72 & $<$0.20       & 0.14 \\
  B0420$-$014 &  469.05 & $<$0.52       & 0.11 \\
  B0838$+$133 &  289.62 & 0.43$\pm$0.07 & 0.08 \\
  B0851$+$202 &  129.86 & $<$0.52       & 0.02 \\
  B1228$+$126 &  ~87.82 & \nodata       & 0.02 \\
  B1663$+$382 &  ~61.79 & $<$0.52       & 0.01 \\
  B2223$-$052 &  241.85 & $<$0.52       & 0.06 \\
  B2249$+$185 &  273.70 & \nodata       & 0.05 \\
  B2251$+$158 &  381.29 & 0.91$\pm$0.04 & 0.09 \\
\tableline
\end{tabular}
\tablenotetext{1}{Integrated \schi\ intensity ($\whi$) obtained from the \lab\ data.}
\tablenotetext{2}{Integrated CO intensity ($\wco$) obtained from previous studies 
	          \citep{liszt1993,liszt2010,li2018}.
		  See Appendix for details.}
\tablenotetext{3}{Total reddening ($\ebv$) from \citet{schlafly2011}.}
\end{table}

\begin{table*}[t]
\small
\setlength{\tabcolsep}{0.065in}
\caption{Column Densities of Hydrogen Atomic and Molecular Gas \label{tab:cN}}
\begin{tabular}{@{}lcccc ccccc@{}}
\tableline
 && \schi && \multicolumn{2}{c}{\h2} && \multicolumn{3}{c}{Total H (\schi\ + 2\h2)} \\  
 \cline{3-3} \cline{5-6} \cline{8-10}
\diagbox[width=6em]{Source}{Tracer} && \schi  && CO & \hco && \schi\ \& CO & \schi\ \& \hco & $\ebv$ \\ 
\tableline
 B0316$+$413 &&    12.9 &&   $<$\,0.4 &   $<$\,0.08&& $<$\,13.7\phn &   $<$\,13.1\phn &    13.5 \\
 B0420$-$014 && \phn8.5 &&   $<$\,1.0 &   $<$\,1.1 && $<$\,10.5\phn &   $<$\,10.7\phn &    10.6 \\
 B0838$+$133 && \phn5.3 &&\phn\phn0.9 &\phn\phn2.7 &&\phn\phn7.1    &    \phn10.7     & \phn7.7 \\
 B0851$+$202 && \phn2.4 &&   $<$\,1.0 &   $<$\,0.4 &&  $<$\,4.4     &    $<$\,3.2     & \phn1.9 \\
 B1228$+$126 && \phn1.6 &&  ~~\nodata &   $<$\,0.3 && ~~\nodata     &    $<$\,2.2     & \phn1.9 \\
 B1663$+$382 && \phn1.1 &&   $<$\,1.0 &   $<$\,0.8 &&  $<$\,3.1     &    $<$\,2.7     & \phn1.0 \\
 B2223$-$052 && \phn4.4 &&   $<$\,1.0 &   $<$\,1.7 &&  $<$\,6.4     &    $<$\,7.8     & \phn5.8 \\
 B2249$+$185 && \phn5.0 &&  ~~\nodata & ~~\nodata  && ~~\nodata     &  ~~\nodata      & \phn4.8 \\
 B2251$+$158 && \phn6.9 &&\phn\phn1.8 &\phn\phn1.2 &&  \phn10.5     & \phn\phn9.3     & \phn8.7 \\
\tableline
\end{tabular}
\tablecomments{Column densities in $10^{20}$ (cm$^{-2}$).
              For the detail description of table entries, see Section~\ref{sec:disc}.
              }
\end{table*}

%
%
%

\subsection{$\ntot$ derived from $\ebv$}
\label{sec:nh_ebv}

$\ebv$ toward each source is obtained from the datacube of \citet{schlafly2011}
which originates from the work of \citet{schlegel1998}.
\citet{schlegel1998} derived $\ebv$ from far-infrared dust emission
at 2\farcm5-pixels with an angular resolution of 6\arcmin\ 
and a measurement error of 16\%.
After that, \citet{schlafly2011} re-examined the values of \citet{schlegel1998}
and provided new estimates, which are somewhat lower (14\% downward) than the original data.
We finally picked the mean value of \citet{schlafly2011} 
for a 5\arcmin-radius circle, with each center provided on the webpage\footnotemark[3].
(See the values listed in the last column of Table~\ref{tab:otherdata}.)
A canonical conversion factor of the dust-to-gas ratio is
$5.8\times10^{21}\,{\rm H}$~(cm$^{-2}$/mag) \citep{savage1977, bohlin1978}.
Recently, however, \citet{liszt2014} examined
the relationship between $\ebv$ and $\nhi$ 
using \schi\ measurements at high latitudes ($|b| \gtrsim 20\arcdeg$),
where neutral atomic gas is very likely to predominate.
\citet{liszt2014} found that
the conversion factor should be higher at $\ebv~\lesssim\,0.1$~mag.
It is $8.3\times10^{21}\,{\rm H}$~(cm$^{-2}$/mag).
Since they used the pre-update $\ebv$ data of \citet{schlegel1998},
we divide by 0.86 to adjust the factor and obtain the equation of
\begin{equation}
\begin{aligned}[b]
\ntot/\ebv = 9.65\times&10^{21}\,{\rm H}~({\rm cm}^{-2}/{\rm mag}) \\ 
 & {\rm for}~\ebv~\lesssim\,0.1~{\rm mag}.
\label{eq:ebv-ntot}
\end{aligned}
\end{equation}
Most LOSs have $\ebv < 0.1$~mag,
while toward B0316+413 and B0420$-$014 are $\ebv=0.14$ and 0.11~mag, respectively. 
We adopted the equation~(\ref{eq:ebv-ntot}) for our all sources 
and obtained values of $\ntot$ 
written in the last column of Table~\ref{tab:cN}.

Comparing between our \hco\ observational results and $\ebv$,
it is interesting that \hco\ has not been detected toward B0316+413 and B0420$-$014 
although their $\ebv$ values are relatively high 
($> 0.1$~mag) compared to the \hco-detected sources.
We also note that their $\ebv$ values are lower or comparable to 
the threshold of \citet{planck2011a19} which is mentioned in Section~\ref{sec:intro}.
So far, \hco\ absorption observations toward 31 LOSs (not counting B2249+185)
at $|b| > 10\arcdeg$ were made by this work and previous studies 
\citep[LL96;][See Appendix for the compiled dataset.]{liszt2000, liszt2010},
and a total of four LOSs (including B1908$-$201 and B1749+096) are in such a case
and also have no CO emission.
If the LOSs have no Galactic molecular gas even DMG, 
is there a possibility of an additional source, such as high-velocity clouds (HVCs), 
increasing $\ebv$?
We checked works of literature and also an \schi\ line profile of \lab\ data,
there seems no HVC toward all LOSs except B1749+096.
\schi\ gas at high velocities ($\vlsr \sim 112-140$~\kms) in the LOS of B1749+096
was reported in \citet{lockman2002}
and suggested to be associated with HVC Complex~C.
However, the presence of dust in Complex~C is controversial 
\citep[e.g.,][]{miville2005, peek2009}.

\subsection{Comprehensive analysis}
 
\begin{figure*}[ht!]
\begin{center}
\includegraphics[width=129mm]{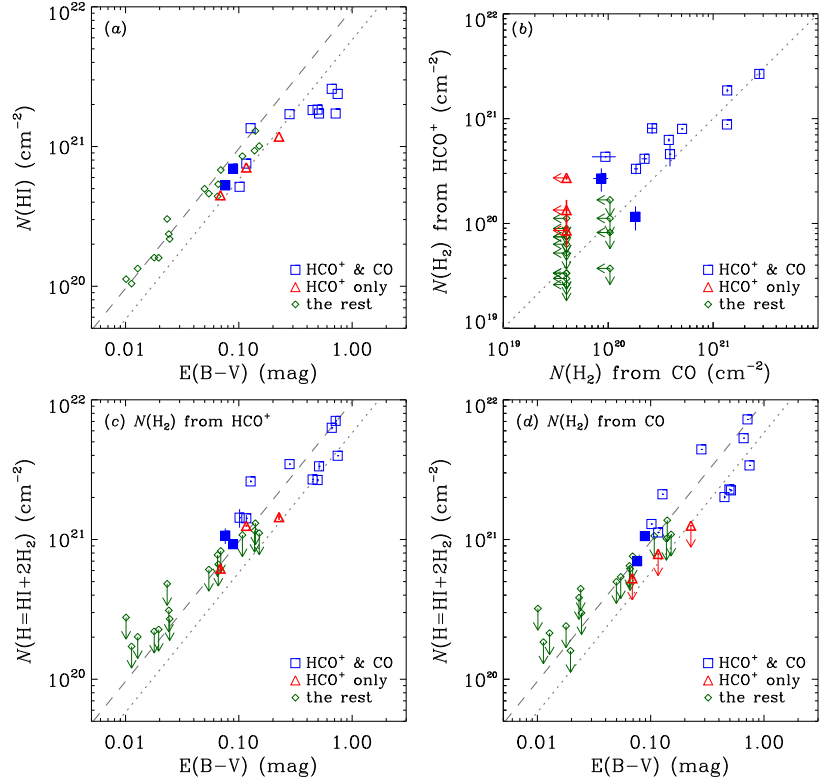}
\caption{
($a$) Column densities of atomic hydrogen vs. total reddening.
Each point is calculated using data listed in Table~\ref{tab:compile} 
and conversion factors described in the text.
Blue squares indicate sources with both \hco\ and CO,
and red triangles are sources with \hco\ but no CO.
Our two targets (B0838+133 and B2251+158) detected in the \hco\ absorption 
are highlighted with a filled symbol.
Green diamonds indicate the other sources without either molecules,
including B2249+185 and B1228+126.
Note that only the available data in the x/y-ranges of each panel are shown.
The accuracy of $\nhi$ obtained from the LAB data is 
about $10^{19}\,{\rm cm}^{-2}$ \citep{kalberla2005}, 
which is negligible compared with the symbol size.
The dashed line is for
$\ntot = 9.65\times10^{21} \ebv ({\rm cm}^{-2}/{\rm mag})$,
while the dotted line is for
$\ntot = 5.8\times10^{21} \ebv ({\rm cm}^{-2}/{\rm mag})$.
($b$)
Comparison of \h2\ column densities derived from two different tracers.
Utilized data points are same as those in ($a$).
Errors of $\nh2$ from \hco\ or CO consider only uncertainties of 
given observational data with fixed factors.
The dotted line is where values of $\nh2$ from \hco\ and CO are equal.
($c$)
Same as ($a$) but total column densities of atomic and molecular hydrogen gas.
$\nh2$ values are obtained from \hco.
($d$)
Same as ($c$) but $\nh2$ from CO.
}
\label{fig:plots_cN}
\end{center}
\end{figure*}

Most \hco-undetected sources do not show CO emission, either.
Their values of $\nhi$ and $\ntot$ from $\ebv$ seem to be consistent,
which suggests that such LOSs are mainly filled with purely atomic gas.
The first panel of Fig.~\ref{fig:plots_cN} shows
a diagram comparing the \schi\ column densities with total reddening.
The data observed in the LOSs at high latitudes ($|b| > 10\arcdeg$),
listed in Table~E1 of \citet{liszt2010} as well as this paper,
are used (See Appendix).
Their column densities are derived using the same ways in this paper.
Green diamonds indicate the LOSs in which neither \hco\ nor CO are seen.
Such sources are well located near a dashed line
which is drawn using a higher conversion factor of $\ntot/\ebv$
than a conventional one (See Section~\ref{sec:nh_ebv}).
It may be hard to constrain the threshold $\ebv$ value
of the \schi-to-\h2\ transition from our results,
but at least any source with $\ebv \lesssim 0.06$~mag might not be DMG.
This result agrees well with the estimate of \citet{liszt2014},
$\ebv \lesssim 0.07$~mag
(The original value of $\ebv$ has been corrected because of the same reason mentioned 
in Section~\ref{sec:nh_ebv}.).

According to the previous studies \citep[e.g., LL96;][]{liszt2012},
most of \hco-detections are within $b \simeq \pm15\arcdeg$,
so the non-detection results from our observations 
(all except one at $|b| > 30\arcdeg$)
are not very surprising.
On the other hand,
considering this work and previous studies together,
six LOSs at $|b| > 15$\arcdeg\ showed \hco-detection.
Half of them, however, show CO-detection:
B0838+133, B2251+158, and B0954+658 ($l, b = 145.746\arcdeg, +43.132\arcdeg$).

Fig.~\ref{fig:plots_cN}b
compares $\nh2$ obtained from the two \h2\ tracers of \hco\ and CO.
At $\nh2 < 10^{21}~{\rm cm}^{-2}$, all \hco-detected sources except B2251+158 
have higher values of $\nh2$ from \hco\ than those from CO.
Figs.~\ref{fig:plots_cN}c--\ref{fig:plots_cN}d
show diagrams of $\ntot$ with respect to total reddening:
the former is $\nh2$ derived from \hco\ and the latter is $\nh2$ derived from CO.
There is a clear difference between the results of the \h2\ tracers at low $\nh2$ and $\ebv$.
Among the fifteen \hco-detected sources, three are not traced by CO.
These are very likely to be dark molecular gas,
and in the range of $0.07 \lesssim \ebv \lesssim 0.2$ or at $\ntot \lesssim 10^{21}~{\rm cm}^{-2}$.
It seems to be shown in panels $c$--$d$ that,
$\ntot$ from \hco\ is systematically larger than
the canonical relation (dotted line);
the relation between $\nh2$ from \hco\ or CO and that from $\ebv$ 
is better described by Equation~(\ref{eq:ebv-ntot}) (dashed line),
even for $\ebv > 0.1$~mag.
Also, the distribution of $\nh2$ derived from \hco\ with respect to $\ebv$
is less dispersed than that derived from CO.
In addition, almost two-thirds of the \hco-detected sources give
larger molecular gas fractions ($f_{\rm H_2} = 2\nh2/\ntot$) than
the typical value of 0.35 \citep[e.g.,][]{liszt2010}.

Finally, 
our two \hco-detected sources, B0838+133 and B2251+158, have similar values of $\ebv$
($\sim 0.1$~mag),
which are within the range shown where it is likely to be DMG.
Although both are traced by CO,
there is a difference between the values of $\nh2$ derived from \hco\ and CO,
as shown in Table~\ref{tab:cN} and Fig.~\ref{fig:plots_cN}b.
That is, the LOS of B0838+133 is expected to have additional amount of gas not traced by CO,
which suggests that the LOS may contain DMG.
However, DMG is not likely to exist toward B2251+158.
Further studies with future observations over a larger region will uncover more details.

\section{Summary}
\label{sec:sum}

We observed nine LOSs of extragalactic compact millimeter wave continuum sources
in \hco~$J=1-0$ absorption line using the KVN 21-m telescope in single dish mode.
Seven of the LOSs were first observed, 
although B2249+185 itself was not seen during our observations.
We detected \hco\ absorption lines in two (B0838+133 and B2251+158)
among the eight LOSs.
The detection toward B0838+133 is a new discovery.
We derived the hydrogen column densities or their limits and
compared them to those inferred from CO line and far-infrared dust continuum emission.
Also, we collected data for other LOSs from the literature.
Our main results are as follows:

(1) In the \hco-undetected LOSs,
CO line emission was not detected, either,
and the values of $\ebv$ are $< 0.1$~mag.
The LOSs are expected to be almost entirely filled with pure atomic gas.
Hydrogen column densities derived from \schi\ line data
are linearly correlated  with those from the values of $\ebv$,
accepting a higher conversion factor of 
$\ntot/\ebv = 9.65\times10^{21}\,{\rm H}~({\rm cm}^{-2}/{\rm mag})$.

(2) In the two \hco-detected LOSs,
CO line emission was also detected and the values of $\ebv$ are similar,
but the differences between 
the values of $\nh2$ estimated from \hco\ and CO line data 
are quite different.
Our \hco\ observational results suggest
that toward B0838+133 there may be a non-negligible amount of
\h2\ gas not fully traced by CO, i.e., DMG.
On the other hand, it is very likely that no or little DMG exists
toward B2251+158.

(3) 
\hco\ absorption was detected toward 15 sources at $|b| > 10\arcdeg$
and CO emission was not detected toward only 3 of them.
The values of $\ebv$ toward the three are 0.07--0.2~mag and,
at that range, \hco\ absorption observations could be useful to
complement the missing component of molecular gas.

\acknowledgments
We are grateful to the staff of the KVN who helped to operate the telescopes. 
The KVN is a facility operated by the KASI (Korea Astronomy and Space Science Institute). 
The KVN observations are supported through the high-speed network connections 
among the KVN sites provided by the KREONET (Korea Research Environment Open NETwork), 
which is managed and operated by the KISTI (Korea Institute of Science and Technology Information).

\nocite{*}
\bibliographystyle{spr-mp-nameyear-cnd}

{}

\appendix
\section{Appendix material}
\label{sec:app_data}

The data compiled in Table~\ref{tab:compile}
are for sight lines of background sources at high latitudes ($|b|>10\arcdeg$)
listed in this paper and Table~E1 of \citet{liszt2010}:
total reddening ($\ebv$),
integrated \schi\ intensity over LSR velocities ($\whi$),
integrated optical depth of \hco\ ($\int{\tau_{HCO^+}}\,dv$),
and integrated CO intensity obtained ($\wco$).
$\ebv$ and $\whi$
are obtained using the same method described
in Sections~\ref{sec:nh_ebv} and \ref{sec:nh_conv}, respectively.
We note that the values in Table~\ref{tab:compile}
give about 14\% lower values than those listed by \citet{liszt2010},
who referred to pre-update data.
For $\int{\tau_{HCO^+}}\,dv$,
we referred to Table~\ref{tab:gaussfit} 
or to \citet[][and references therein]{liszt2010}.
For a common line of sight between the two sets of data,
the value from Table~\ref{tab:gaussfit} was retained.
And, $\wco$ values were taken from previous studies.
We have adopted the values of $\wco$ provided by \citet{liszt2010}, if available.
Otherwise, we have referred to the data of \citet{liszt1993} or \citet{li2018}.
No targets with $\wco$ from \citet{liszt1993} were detected
in the CO emission, so we give an upper limit assuming one Gaussian component
with the peak temperature (nominal sensitivity limit; $\sim 0.33$~K in main beam scale)
and a line width of 1.5~\kms.

\setcounter{table}{0}
\renewcommand{\thetable}{A\arabic{table}}
\begin{table*}[ht!]
\small
\caption{Radio Continuum Sources at $|b|>10\arcdeg$ with \hco\ Absorption Measurement \label{tab:compile}}
\begin{tabular}{@{}rlcccrcc@{}}
\tableline
\# & Source & $\ell$ & $b$ & 
$\ebv$ & $\whi$ &
$\int{\tau_{\rm HCO^+}}\,dv$ & $\wco$\tablenotemark{*} \\ 
 & & (\arcdeg) & (\arcdeg) & 
(mag) & (\Kkms) & (\kms) & (\Kkms) \\
\tableline
1  & B1730$-$130 & ~12.032 & $+$10.812 & 0.4478 & 1002.97 & 1.16$\pm$0.02 & 0.47$\pm$0.12 \\
2  & B1908$-$201 & ~16.857 & $-$13.219 & 0.1376 &  513.53 & $<$0.30       & $<$0.20       \\
3  & B1741$-$038 & ~21.591 & $+$13.128 & 0.4955 & 1009.97 & 1.11$\pm$0.10 & 1.11$\pm$0.07 \\
4  & B1749$+$096 & ~34.920 & $+$17.644 & 0.1514 &  553.89 & $<$0.14       & $<$0.20       \\
5  & B2223$-$052 & ~58.960 & $-$48.843 & 0.0649 &  241.85 & $<$0.45       & $<$0.52\tablenotemark{**}       \\
6  & B1663$+$382 & ~61.085 & $+$42.337 & 0.0101 &   61.79 & $<$0.22       & $<$0.52\tablenotemark{**}       \\
7  & B1641$+$399 & ~63.455 & $+$40.948 & 0.0113 &   57.40 & $<$0.09       & $<$0.20       \\
8  & B2145$+$067 & ~63.656 & $-$34.072 & 0.0688 &  246.25 & 0.23$\pm$0.07 & $<$0.20       \\
9  & B1954$+$513 & ~85.298 & $+$11.757 & 0.1270 &  743.34 & 1.68$\pm$0.06 & 1.89$\pm$0.04 \\
10 & B1823$+$568 & ~85.739 & $+$26.080 & 0.0544 &  252.92 & $<$0.20       & $<$0.20       \\
11 & B2251$+$158 & ~86.111 & $-$38.184 & 0.0889 &  381.29 & 0.31$\pm$0.08\tablenotemark{\dag} & 0.91$\pm$0.04\tablenotemark{\ddag} \\
12 & B2249$+$185 & ~87.354 & $-$35.648 & 0.0498 &  273.71 & \nodata       & \nodata       \\
13 & B2200$+$420 & ~92.590 & $-$10.441 & 0.2802 &  936.76 & 2.36$\pm$0.03 & 6.78$\pm$0.05 \\
14 & B1928$+$738 & 105.625 & $+$23.541 & 0.1163 &  389.30 & 0.73$\pm$0.03 & $<$0.20       \\
15 & B0212$+$735 & 128.927 & $+$11.964 & 0.6608 & 1421.96 & 4.98$\pm$0.20 & 6.81$\pm$0.06 \\
16 & B0954$+$658 & 145.746 & $+$43.132 & 0.1015 &  282.94 & 1.23$\pm$0.29 & 1.94$\pm$0.04 \\
17 & B0316$+$413 & 150.576 & $-$13.261 & 0.1399 &  710.72 & $<$0.02       & $<$0.20       \\
18 & B0235$+$164 & 156.772 & $-$39.108 & 0.0693 &  373.42 & $<$0.20       & $<$0.20       \\
19 & B0923$+$392 & 183.709 & $+$46.165 & 0.0128 &   73.57 & $<$0.09       & $<$0.20       \\
20 & B0528$+$134 & 191.368 & $-$11.012 & 0.7457 & 1307.63 & 2.14$\pm$0.02 & 2.53$\pm$0.06 \\
21 & B0420$-$014 & 195.291 & $-$33.140 & 0.1075 &  469.05 & $<$0.30       & $<$0.52\tablenotemark{**}       \\
22 & B0851$+$202 & 206.812 & $+$35.821 & 0.0241 &  129.86 & $<$0.10       & $<$0.52\tablenotemark{**}       \\
23 & B0838$+$133 & 212.968 & $+$30.139 & 0.0759 &  289.62 & 0.72$\pm$0.18 & \phn\phn0.43$\pm$0.07\tablenotemark{***} \\
24 & J0008$+$686 & 215.752 & $-$13.253 & 0.7155 &  947.02 & 7.16$\pm$0.75 &13.79$\pm$0.05 \\
25 & B0605$-$085 & 215.752 & $-$13.523 & 0.5105 &  947.02 & 2.17$\pm$0.25 & 1.31$\pm$0.13 \\
26 & B0736$+$017 & 216.990 & $+$11.380 & 0.1164 &  414.70 & 0.89$\pm$0.10 & 0.92$\pm$0.04 \\
27 & B0607$-$157 & 222.611 & $-$16.183 & 0.2263 &  647.22 & 0.36$\pm$0.09 & $<$0.20       \\
28 & B1055$+$018 & 251.513 & $+$52.775 & 0.0233 &  166.52 & $<$0.24       & $<$0.20       \\
29 & B1228$+$126 & 283.778 & $+$74.491 & 0.0196 &   87.82 & $<$0.09       & \nodata       \\
30 &       3C273 & 289.954 & $+$64.360 & 0.0179 &   88.11 & $<$0.08       & $<$0.20       \\
31 &       3C279 & 305.107 & $+$57.062 & 0.0245 &  119.90 & $<$0.07       & $<$0.20       \\
32 & B1334$-$127 & 320.026 & $+$48.374 & 0.0658 &  295.29 & $<$0.17       & $<$0.20       \\
\tableline
\end{tabular}
\tablenotetext{*}{
		  Values without an alphabet letter are taken from Table~E1 of \citet{liszt2010};
		  values with ** are  from \citet{liszt1993} and 
                  values with *** are from \citet{li2018}.}
\tablenotetext{\dag}{This was reported in previous research papers of \citet{lucas1996} and \citet{liszt2000}.
		  The recent result gives a value of 0.280$\pm$0.005~\kms\ using single-component Gaussian fit, 
		  which seems to be consistent with our given value within $1\sigma$ uncertainty.
                 }
\tablenotetext{\ddag}{\citet{li2018} gives a slightly lower value, $\wco=0.76\pm0.08$~\Kkms.}
\end{table*}

\end{document}